\documentstyle[epsfig]{mn}

\title[SBF distances to Leo and Virgo using the HST ]{SBF distances  to Leo and
Virgo using the HST\thanks{Based on observations made with the NASA/ESA Hubble
Space Telescope, obtained from the data archive at the Space Telescope Science
Institute. STScI is operated by the Association of Universities for Research in
Astronomy, Inc. under the NASA contract NAS 5-26555.}.} 

\author[Morris \& Shanks]
{P.W. Morris$^1$ and T. Shanks$^1$ \\
$^1$ Department of Physics, Durham University, South Road, Durham, DH1 3LE.
}

\date{Accepted-     \hspace*{2cm}   . Received \hspace*{2cm}   ; in original form}

\begin{document}
\newcommand{\ie}{{\it i.e. }}
\newcommand{\etal} {{\it et al} }
\newcommand{\DM}{$(m-M)_0$}
\setcounter{figure}{0}
\setcounter{table}{0}

\maketitle

\begin{abstract}
   
We have used archive HST  WFPC2 data for three elliptical galaxies (NGC
3379 in the Leo I group, and NGC 4472 and  NGC 4406 in the Virgo
cluster) to determine their distances using the Surface Brightness
Fluctuation (SBF) method as described by Tonry and Schneider (1988). A
comparison of the HST results with the SBF distance moduli of
Ciardullo et al (1993) shows significant disagreement and suggests
that the r.m.s. error on these ground-based distance moduli is
actually as large as $\pm$0.25 mag.  The agreement is only slightly
improved when we compare our results with the HST and ground-based SBF
distances from Ajhar \etal (1997) and Tonry  \etal (1997); the
comparison suggests that a lower limit on the error of the HST SBF
distance moduli is $\pm$0.17 mag. Overall, these results suggest that
previously quoted measurement errors may underestimate the true error
in SBF distance moduli by at least a factor of 2-3.
   
\end{abstract}


\section{Introduction}

There now exist several methods of distance determination that appear to
give measurements with small errors, \ie the SBF method and the use of
planetary nebula luminosity function (PNLF). Both of theses methods claim
(and appear to deliver) distances with an accuracy of better than 10\%.
These methods are crucial in determining the zero-point and
reliability of other distance estimators and also in determining
the structure of the Universe in our neighbourhood out to about 30Mpc.

In this paper we present an analysis of archive HST WFPC2 data of three
ellipticals using the SBF method. The SBF method is based on the
pixel-to-pixel brightness variations that are present in all galaxies,
which is due to the varying number of stars within each pixel. The
variations are proportional to $\overline{f}\sqrt{N_*}$ where
$\overline{f}$ is the mean flux of a star and $N_*$ is the average
number of stars per pixel. To detect the SBF signal, it must be strong
enough to be separated from sources of noise within the frame. The main
sources of noise are shot noise and contamination from faint objects
within the frame. The shot noise can be overcome by taking a long
enough exposure, but the contamination from point sources needs better
resolution to reduce its effect. Since the resolution of the HST is far
superior to anything yet possible from the ground then images taken
with the HST should reduce the point source contamination
considerably. The HST data therefore will provide a crucial check of
the accuracy of the ground-based SBF technique.

\begin{table}
   \begin{tabular}{|c|c|c|c|c|} \hline
      Galaxy & P.I. & Proposal ID & F814W & F555W \\ \hline
      NGC 3379 & Faber & 5512 & 3x500,1x160 & 3x400,1x140 \\
      NGC 4472 & Westphal & 5236 & 2x900 & 2x900 \\
      NGC 4406 & Faber & 5512 & 3x500 & 3x500 \\ \hline
   \end{tabular}
   \caption{HST archive data used for the SBF analysis. Exposure times for
   each filter are given in seconds.} \label{ta:archive}
\end{table}

\section {The Observations}

A search of the HST archive provided us with images of three galaxies, on
which to perform the SBF analysis, with previous ground-based SBF distances
which can then be checked. WFPC2 frames taken using the F814W and F555W
filters were available for each of the galaxies NGC 3379, NGC 4472 and NGC
4406. The details of the frames are given in Table~\ref{ta:archive}.

All of the archive frames have the nucleus of the galaxy centred on the PC
chip, with the outer regions present in all of the WF chips. On none of the
frames were the central regions of the galaxy saturated.

\section {Data reduction} \label{sec:reduc}
   
All of the data were reduced using software available under
IRAF\footnote{IRAF is distributed by the National Optical Astronomical
Observatories} and some specially written software by the authors.
The frames for each galaxy and filter were co-added and any cosmic rays
removed using the standard tools available under IRAF in the STSDAS
package.

\subsection{Removing the galaxy background}

Before we are able to investigate the SBF, the galaxy background needs to
be removed from the images. 

For the PC images an ellipse fitting routine was used ({\it ellipse} under
the {\it stsdas.analysis.isophote} package). This routine iteratively fits
elliptical isophotes over a 2-dimensional image and produces an output which
can be used to construct a smooth image of the fit. This image was then
subtracted from the original image to give a resultant image which is flat
on large scales. Generally the fitting worked well, but very close into the
nucleus of the galaxy it failed. For some PC images the fitting broke down
completely at the edges of the chip.

For the WF images it was not possible to use any of the ellipse fitting
routines available under IRAF since they all required that the center of the
ellipse was in the image. Instead, we followed Simard \& Pritchet
(1994) and used {\it fit1d} to provide a fit to the galaxy
background. For each frame thirty {\it fit1d} passes were made with five
2.5$\sigma$ rejection iterations for each pass. For all the WF frames this
provided a flat image, except for the bottom $\approx 200$ pixels of each
frame where the subtracted image showed some structure.

\subsection {Detection and removal of point sources} \label{sec:lf}
   
Once a flat image was obtained, DAOPHOT was used to find all sources
greater than 4 sigma above the background and determine their magnitudes. 
To determine the magnitude of the point sources aperture photometry was
used, with an aperture radius of $0.''5$. Crowding was not a problem on any of
the frames. After zero-pointing the magnitudes of the point sources using
the method described in Holtzman \etal (1995), a
point source number count was then constructed using 0.5mag
bins down to the completeness limit of the data.

\begin{figure*}
   \epsfig{file=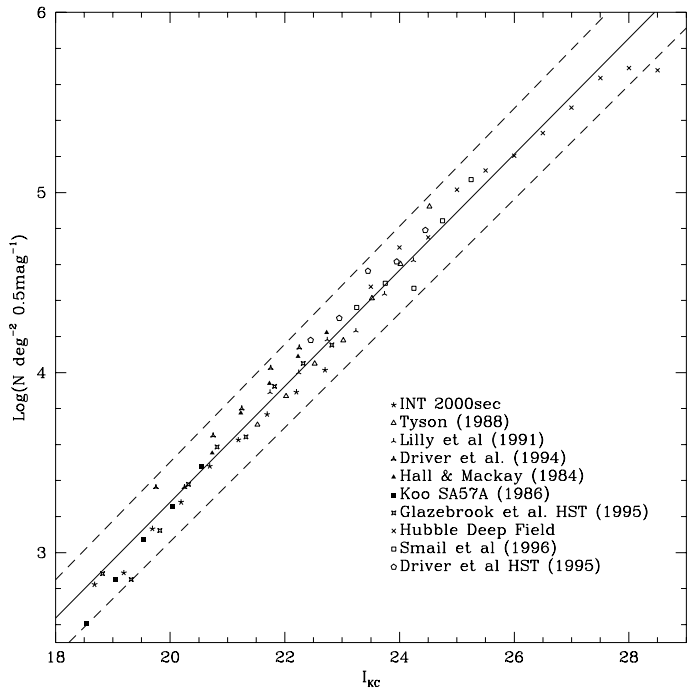,height=10.0cm}
   \caption{Galaxy counts taken from a range of sources. All magnitudes have been 
   converted to $I_{KC}$ and have been de-reddened. The straight line is the best least
   squares fit to all of the data points. The dashed lines show the error
   limit of the fit. } \label{fi:backgals}
\end{figure*}

The sum of a Gaussian globular cluster luminosity function (GCLF)
\cite{Harris88} and a power-law number count for background galaxies
was then fitted to the observed point source number count. For the GCLF
$\sigma = 1.4$ mag was adopted and $M^V_0 = -7.4$ \cite{Harris88}. For
the background galaxies the slope and zero-point of the number count
was found by combining I band counts from a variety of sources
\cite{Tyson88,Lilly91,Driver94,HM84,Koo86,Glaze95,Smail96,Driver95} and
then fitting a power law to the combined counts (see
Figure~\ref{fi:backgals}).  This gave a slope of $0.322\pm0.005$ for
the I band counts, with an intercept of $-3.158 \pm 0.122$. The galaxy
number count was fixed, as was the apparent magnitude of the GCLF,
while the amplitude of the GCLF was allowed to vary to provide the best
fit. The apparent magnitude of the peak of the GCLF was first set by
the distances given in Ciardullo \etal (1993) for each
galaxy, then determining the distance to the galaxy using the HST data,
recalculating the peak based on the new distances, and then iterating.

Once the best fit to the point source number count was determined it
was then possible to calculate the contribution to the variance in the image
that point sources beyond our cutoff magnitude ($I_{KC}\sim$24mag) make. 
This was done by integrating the fitted number count past the cutoff, using
the equation for the residual power $P_r$ 

\begin{equation}
P_r = \sum n(m)f^2(m)/\overline{g}
 \label{eqn:pr}
\end{equation}

\noindent where $n(m)$ is number of undetected point sources per pixel in
the magnitude bin centred on $m$, $f(m)$ is the flux of a source of
magnitude $m$ and $\overline{g}$ is the average galaxy flux per pixel.

\subsection{Determining the SBF power by Fourier analysis}

To determine the power present in the image due to SBF signal it is
necessary to construct a Fourier power spectrum of the image. Since the SBF
signal has the effect of the point spread function (PSF) of the
telescope/detector imprinted on it, it is possible to separate it from the
other sources of variance (shot noise, readout noise) that appear as white
noise in the power spectrum (\ie equal power at all scales).

Before the power spectrum is found for the image, a mask image is
constructed. The mask image is 1 everywhere except for areas of 20 by 20
pixels, which are set to zero, centred on the positions of all point
sources down to the magnitude limit used above to calculate the number count.
Also, any defects and, in the case of the PC chips, the central regions of the
galaxy were masked out.

The image was then multiplied by the mask, and the observed Fourier power
spectrum ($P(k)$) calculated for the masked image, using the {\it
powerspec} task available under STSDAS/IRAF. The output power spectrum was
normalized by the mean galaxy flux of the unmasked pixels in the image.

An expectation power spectrum $E(k)$ was constructed for each region by
convolving the power spectrum of the PSF  with the power
spectrum of the mask for that region. The PSF for each region was determined
by using a grid of artificial PSFs provided by Tanvir (private
communication), who used the {\it Tiny Tim} \cite{Kirst94} program to
calculate artificial PSFs. In regions where bright stars were present the
power spectra of the stars were found and compared to the power spectrum
from the artificial PSFs. The agreement was found to be extremely good. The
expectation power spectrum were normalised so that $E(k=0) = 1$. For each of
the images under study the observed power spectrum was fitted using the
relationship

\begin{equation}
P(k) = P_0 E(k) + P_1
\label{equation1}
\end{equation}

\noindent where $P_0$ is the sum of the the power in the image due to SBF
and residual point sources, \ie $P_0 = P_{fluc} + P_r$

\begin{figure*}
   \epsfig{file=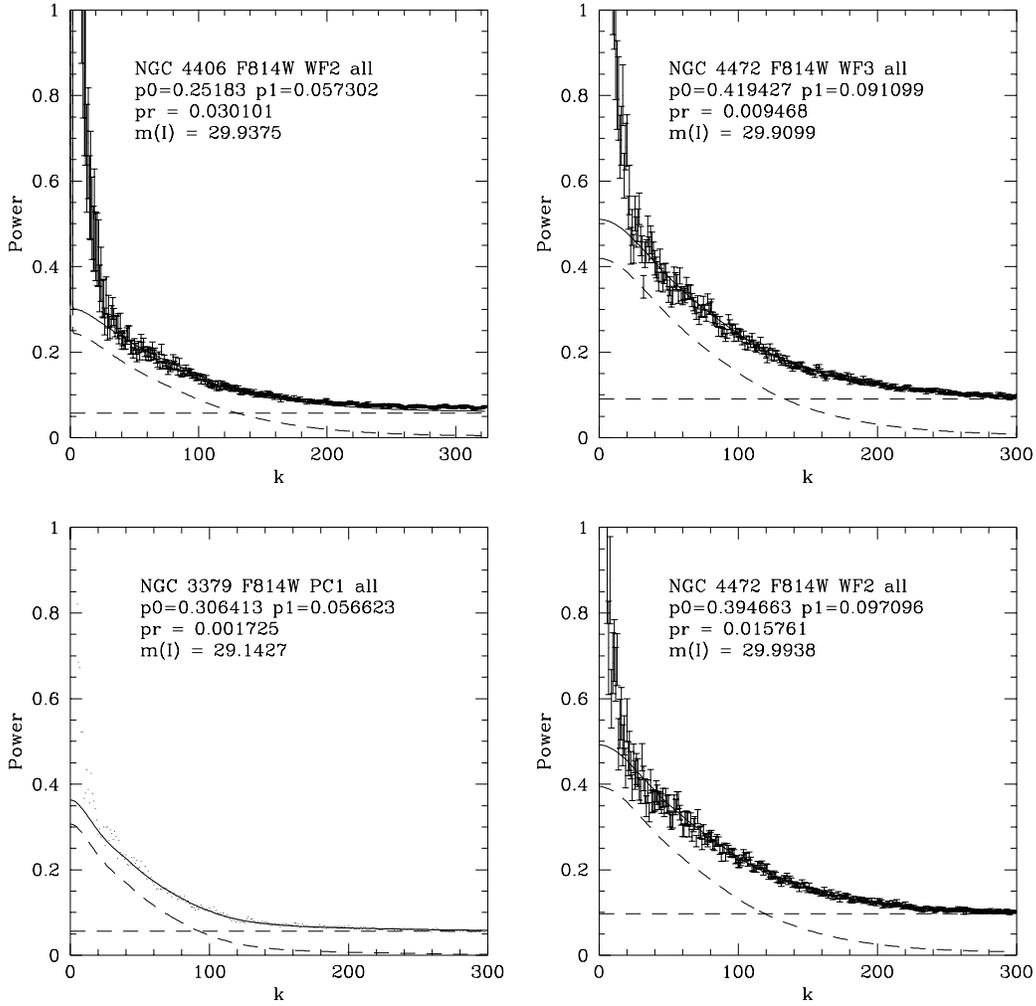,height=15.0cm}
   \caption{The observed power spectra and fitted PSF power spectra for four
   HST frames. The points represent the observed power spectra, the straight
   dashed line is $P_1$ and the curved dashed line is $P_0E(k)$, the scaled
   PSF power spectrum. The solid line is the sum of these two, i.e.
   $P_0E(k) + P_1$, where $P_1$ and $P_0$ are the best fit values to the
   data. No error bars are shown for one plot so that the scatter about the
   fitted line can be seen.} \label{fi:powspec}
\end{figure*}

To be able to fit the data using Equation~(\ref{equation1}), the two
dimensional power spectra (as output from the 2D Fourier analysis) were
converted into one dimensional power spectra by azimuthally averaging each
region --- each region was deliberately made square to make this task
easier. The expectation power spectrum was then fitted to $P(k)$ using an
iterative least squares method over several ranges of wave numbers
(typically $40<k<70$,$70<k<100$,$100<k<130$,$130<k<160$,$160<k<190$) and a
weighted average values of the parameters $P_0$ and $P_1$ calculated. This
method was used so that the stability of the fits could be tested over the
range of the power spectrum. Examples of the power spectra obtained and the
fits calculated in the spectra are shown in Figure~\ref{fi:powspec}.

\subsection{Comparison of observed $P_1$ values with theory} \label{se:ct}
   
As a check on the fitting of the power spectrum and also the method we
have used, it is possible to check the observed value of $P_1$ with that
expected from the theory of SBF.

From Tonry \& Scheider \shortcite{TS88} $P_1$ is given in their Equation
23 as

\begin{equation}
P_1 = \sum \sigma^2_R + \sum \sigma^2_p
\end{equation}

where

\begin{equation}
\sigma^2_R = N^2_R / a^2
\end{equation}

and

\begin{equation}
\sigma^2_p = \overline{c}/a
\end{equation}

\noindent where $a$ is the gain in electrons per ADU, $N_R$ is the readout
noise in electrons and $\overline{c}$ is the mean value of the sky plus
galaxy for the image. In the case of WFPC2, $a = 7$ and $N_R = 5$ which
gives $ \sigma^2_R = 0.0104 $ADU. For all the galaxies here, multiple frames
have been averaged together, so that the observed $P_1$ will actually be
$P_1 / \sqrt N_{frames}$ . A comparison of the observed and predicted values
for $P_1$ is given in Table~\ref{ta:p1}. As can be seen from this table
there is good agreement between the theory and observations for the galaxies
under study.

\begin{table}
   \begin{tabular}{|c|c|c|c|} \hline
      Galaxy & Chip & Predicted & Observed \\
      &  &  $P_1$ & $P_1$ \\ \hline
      NGC 4472 & WF2 & 0.103 & 0.095 \\
       & WF3 & 0.080 & 0.080 \\
      NGC 4406 & WF2 & 0.068 & 0.058 \\
      & WF3 & 0.062 & 0.071 \\
      & WF4 & 0.064 & 0.062 \\
      NGC 3379 & WF2 & 0.034 & 0.047 \\
      & WF3 & 0.027 & 0.060 \\ \hline
   \end{tabular}
   \caption{A comparison of the measured and predicted values for $P_1$. All
   values of $P_1$ are given in ADU.}
   \label{ta:p1}
\end{table}

\subsection {The effect of binning the images}

To investigate whether the pixel size had any effect on the results
obtained, we took several images and binned them by various factors and
then performed the complete SBF analysis on the binned images. The results are shown
in Table~\ref{ta:bin} and show that this does not significantly change our results.

\begin{table}
   \begin{tabular}{|c|c|c|c|} \hline
      Galaxy & Chip & Bin Size & $(M-m)_0$ \\ \hline
      NGC 4472 & WF2 & 1x1 & 31.01 \\
      & & 2x2 & 31.06 \\
      & & 4x4 & 31.27 \\ \hline
      NGC 3379 & WF2 & 1x1 & 30.28 \\
      & & 2x2 & 30.37 \\
      & & 3x3 & 30.33 \\ \hline
   \end{tabular}
   \caption{The effects on $(M-m)_0$ of binning several images by different amounts.}
   \label{ta:bin}
\end{table}

\subsection {Galaxy colours}

The absolute SBF magnitude is strongly dependent on the colour of the galaxy
under study (see Equation~\ref{eqn:MI}), hence the colour needs to be
determined accurately. To determine the colour, (V-I), for the regions
observed we used the fits to the galaxy background in both filters. The fits
were then multiplied by the mask image, and the mean flux found for the
unmasked pixels in the region used. To determine values for the sky
flux, the median value was found in the corner of each frame furthest
from the centre of the galaxy. Then, using Equation (8) and the values in
Table 7 in Holtzman  \etal (1995) and an interactive
method, the standard V, I and (V-I) magnitudes were calculated.

\section{Simulations}

To confirm that the method that was used is valid simulations of all of
the galaxies were performed. For each of the galaxies studied we took the
galaxy fit for the WF2 frame and to this added simulated surface brightness
fluctuations by adding to each pixel in the frame a random Gaussian deviate,
with the variance given by Equation 10 from Tonry \& Schnieder (1988) and reproduced here,

\begin{equation}
\sigma^2_L = \overline{g}(x,y)t{\frac{10 pc}{d}}^2 10^{-0.4(\overline{M}
- m_1)}
\end{equation}

\noindent where $\sigma^2_L$ is the SBF variance, $\overline{g}(x,y)$ is the
galaxy signal in ADU at (x,y), $t$ is the exposure time in seconds, $d$ is
the distance to the galaxy in parsecs, $\overline{M}$ is the absolute
fluctuation magnitude for the galaxy and $m_1$ is the magnitude of an object
that produces 1 ADU $s^{-1}$ on the detector/telescope used. After adding
the SBF variance to the galaxy background, the whole image was then
convolved with a PSF for the image, using the {\it fconvolve} STSDAS IRAF
task. Photon and read noise appropriate for the WFPC2 was then added, to
create the final simulated image. No points sources/ background galaxies were
added to the image.

The simulated frames were then reduced in the same manner as the real
frames, except that no point source number counts were calculated and the mask
images were the same as those used for the real data. For all the simulations
performed the recovered distant modulus agreed with the input distance
modulus to within 0.05 mag.

\section {Error analysis}

The error in the final distance modulus was determined by evaluating the errors at all
of the stages of the reduction and adding the individual errors in quadrature. The
sources and representative values are shown in Table~\ref{ta:errors}.

\begin{table}
   \begin{centering}
   \begin{tabular}{|c|c|} \hline
      Quantity & Approximate Error \\ \hline
      $P_1$ & $\pm 8\%$\\
      $P_R$ & $\pm P_R / 2$ \\
      HST zero-point & $\pm 2\%$ \\
      (V-I) & $\pm 5\%$ \\ \hline
   \end{tabular}
   \caption{Sources of errors in final distance modulus. The error for $P_R$ is
   difficult to estimate, so we have taken it be half the value of $P_R$. The error
   in the HST zero-point has been taken from Holtzman \etal (1995).}
   \label{ta:errors}
   \end{centering}
\end{table}

\section{SBF distances}

Once the colour and $P_{fluc}$ have been determined for each of the images,
it is then possible to determine the distance modulus. The absolute
fluctuation magnitude ($\overline{M_I}$) is a function of the galaxies
colour, and the dependence has been calibrated by Tonry \shortcite
{Tonry91}, with the dependence given by

\begin{equation}
\overline{M_I} = -4.84 + 3(V-I)_0
\label{eqn:MI}
\end{equation}

\begin{figure}
   \epsfig{file=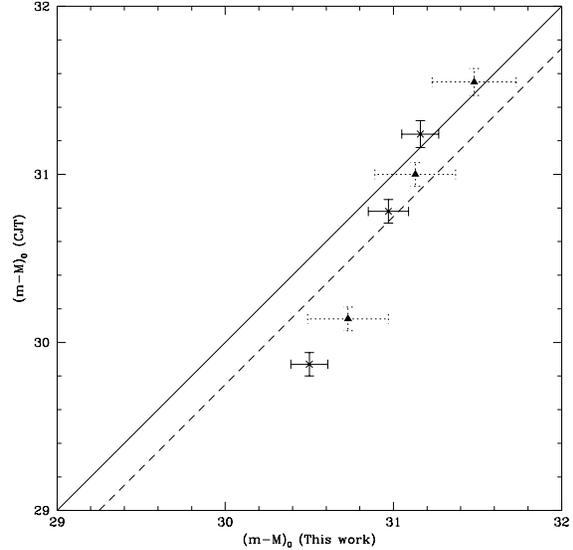,height=8.0cm}
   \caption{A comparison of our results for \DM with those published
   in CJT. The solid line shows the 1:1 relationship, while the dashed line
   is offset by the mean difference between our results and CJT's. The
   observed dispersion around the solid 1:1 line is much bigger than the
   statistical errors would imply. Even around the dashed line, which is
   shifted by 0.25mag to represent the mean difference between CJT and
   ourselves, the observed dispersion is much larger than expected. The triangles
   with the dotted error bars are when the distances are calculated using the
   new calibration given by Tonry \etal (1997).} \label{fi:us_tonry}
\end{figure}

Hence the distance modulus is given by

\begin{equation}
   (m-M)_0 = m_{fluc} + 4.84 - 3(V-I)_{obs} + 0.80A_B
\end{equation}

\noindent where $m_{fluc}$ is the apparent fluctuation magnitude and $A_B$
is the absorption present (this is Equation 3 from Ciardullo \etal
(1993, hereafter referred to as CJT)). For each galaxy we have used the
values of $A_B$ (see Table~\ref{ta:extinc}) given in CJT, which were derived
from Burstein \& Heiles \shortcite{BH84} and  Burstein \etal (1987). The results for each galaxy are given in
Tables~\ref{ta:results} and a comparison of the results for $m_{fluc}$ with
previous ground based results is given in Table~\ref{ta:comp}, and shown
graphically in Figure~\ref{fi:us_tonry}.

\begin{table}
   \begin{tabular}{|c|c|} \hline
      Galaxy & $A_B$ \\ \hline
      NGC 3379 & 0.05 \\
      NGC 4406 & 0.11 \\
      NGC 4472 & 0.00 \\ \hline
   \end{tabular}
   \caption{Foreground extinction values used for each galaxy, taken from
   CJT.}
   \label{ta:extinc}
\end{table}

\begin{table}
   \begin{tabular} {|c|c|c|c|c|}                                                 \hline
      Galaxy & Chip   & $m_{fluc}$       &$(V-I)_{obs}$    & $(m-M)_0$        \\ \hline
      NGC 3379 & PC1  & $29.15 \pm 0.10$ & $1.24 \pm 0.05$ & $30.32 \pm 0.18$ \\
      &          WF2  & $29.27 \pm 0.18$ & $1.29 \pm 0.05$ & $30.28 \pm 0.23$ \\
      &          WF3  & $29.57 \pm 0.28$ & $1.28 \pm 0.05$ & $30.61 \pm 0.32$ \\
      &          WF4  & $29.42 \pm 0.09$ & $1.17 \pm 0.05$ & $30.78 \pm 0.18$ \\ \hline
      &          mean & $29.35 \pm 0.09$ & $1.24 \pm 0.03$ & $30.50 \pm 0.12$ \\ 
      &      weighted & $29.31 \pm 0.06$ &                 & $30.50 \pm 0.11$ \\ \hline

      NGC 4472 & PC1  & $29.92 \pm 0.10$ & $1.26 \pm 0.05$ & $30.99 \pm 0.18$ \\
      &          WF2  & $29.99 \pm 0.14$ & $1.27 \pm 0.05$ & $31.01 \pm 0.21$ \\
      &          WF3  & $29.91 \pm 0.20$ & $1.29 \pm 0.05$ & $30.87 \pm 0.25$ \\ \hline
      &          mean & $29.94 \pm 0.08$ & $1.27 \pm 0.01$ & $30.96 \pm 0.09$ \\ 
      &      weighted & $29.94 \pm 0.08$ &                 & $30.97 \pm 0.12$ \\ \hline
      
      NGC 4406 & PC1  & $29.85 \pm 0.06$ & $1.27 \pm 0.05$ & $30.97 \pm 0.16$ \\
      &          WF2  & $29.94 \pm 0.31$ & $1.21 \pm 0.05$ & $31.24 \pm 0.34$ \\
      &          WF3  & $30.22 \pm 0.12$ & $1.32 \pm 0.05$ & $31.20 \pm 0.19$ \\
      &          WF4  & $30.28 \pm 0.22$ & $1.23 \pm 0.05$ & $31.54 \pm 0.27$ \\ \hline
      &          mean & $30.07 \pm 0.10$ & $1.26 \pm 0.03$ & $31.24 \pm 0.12$ \\ 
      &      weighted & $29.94 \pm 0.05$ &                 & $31.16 \pm 0.11$ \\ \hline
   \end{tabular}
   \caption{Results of the SBF analysis for all three galaxies. $m_{fluc}$ and V-I are
both uncorrected for absorption. The measurement errors on individual frames are estimated
as indicated in Table~\ref{ta:errors}. The errors on the mean are  estimated from the rms
variation between frames in the cases of NGC3379 and NGC4406 and from the average
instrumental errors added in quadrature in the case of NGC4472. The weighted means (and
errors) are calculated using the measurement errors shown for each individual frame.}
   \label{ta:results}
\end{table}
 
\begin{table}
   \begin{tabular} {|c|c|c|r|}                                              \hline
      Galaxy   & $m_{fluc}$       & $m_{fluc}$        & \multicolumn{1}{c}{diff}\\ 
               &  (this)          & (CJT)             &            \\ \hline
      NGC 3379 & $29.31 \pm 0.06$ & $ 28.65 \pm 0.07$ & $ 0.66 \pm 0.09$ \\
      NGC 4472 & $29.94 \pm 0.08$ & $ 29.66 \pm 0.07$ & $ 0.28 \pm 0.11$ \\
      NGC 4406 & $29.94 \pm 0.05$ & $ 29.97 \pm 0.07$ & $-0.03 \pm 0.09$ \\ \hline
      Galaxy   & $(m-M)_0$        & $(m-M)_0$         & \multicolumn{1}{c}{diff}\\
               &  (this/CJT)      &   (CJT)           &                  \\ \hline
      NGC 3379 & $30.50 \pm 0.11$ & $ 29.87 \pm 0.07$ & $ 0.63 \pm 0.13$ \\
      NGC 4472 & $30.97 \pm 0.12$ & $ 30.78 \pm 0.07$ & $ 0.19 \pm 0.14$ \\
      NGC 4406 & $31.16 \pm 0.11$ & $ 31.24 \pm 0.08$ & $-0.08 \pm 0.14$ \\
      \multicolumn{3}{c}{Mean offsets}                & $ 0.25 \pm 0.21$ \\
      \multicolumn{3}{c}{r.m.s. error}                & $      \pm\, 0.36$ \\ \hline
      Galaxy   & $(m-M)_0$        & $(m-M)_0$         & \multicolumn{1}{c}{diff}\\ 
               &  (this/TBAD)     &   (CJT)           &                  \\ \hline
      NGC 3379 & $30.73 \pm 0.24$ & $ 30.14 \pm 0.07$ & $ 0.59 \pm 0.25$ \\
      NGC 4472 & $31.13 \pm 0.24$ & $ 31.00 \pm 0.07$ & $ 0.13 \pm 0.25$ \\
      NGC 4406 & $31.48 \pm 0.25$ & $ 31.55 \pm 0.08$ & $-0.07 \pm 0.26$ \\
      \multicolumn{3}{c}{Mean offsets}                & $ 0.22 \pm 0.20$ \\ 
      \multicolumn{3}{c}{r.m.s. error}                & $      \pm\, 0.35$ \\ \hline
   \end{tabular}

\caption{A comparison of the SBF fluctuation magnitudes and distance moduli (assuming the
$M_I:V-I$ calibration of CJT) presented here and those previously determined from ground
based observations taken from CJT. $m_{fluc}$  is uncorrected for absorption. The final
three rows are the values for $(m-M)_0$ using the new calibration of $M_I$ with (V-I) given
by Tonry \etal (1997, denoted by TBAD) and with slope -4.5. For individual galaxies,
weighted means and errors for all quantities are quoted. The mean offsets and
errors are unweighted, with the error here being on the mean and the figure
in brackets being the error on an individual galaxy offset.}
   \label{ta:comp}
\end{table}

\section{Comparison with SBF results of Ciardullo et al, 1993}

By comparing the unweighted and weighted errors in  Table~\ref{ta:results}, we
see that the r.m.s. spread in the values of $m_{fluc}$ and $(m-M)_0$ for our
observations is  consistent with the calculated errors from the fitting  of the
power-spectrum and measuring V-I. The average (weighted) final values of $(m-M)_0$ for
each  galaxy have an error of order  $\pm$0.12 mag, indicating that our HST SBF measures 
provide distances with reasonable internal accuracy.

Table~\ref{ta:comp} shows that there is a significant, 4.8$\sigma$,
disagreement between our result for the distance modulus of NGC 3379
and that of CJT. There is a less significant disagreement over the
distance of NGC 4472 and the distance for NGC 4406 is consistent with
that given by CJT. Table~\ref{ta:comp} and Figure~\ref{fi:us_tonry}
shows that, on average, the CJT distances are about 12\% less than
ours. However, given  at least  our error estimates for the HST SBF
method, this offset is insignificant. Table~\ref{ta:comp} and
Figure~\ref{fi:us_tonry} also show that the scatter around the mean
offset, (dashed) line is larger than that expected from the claimed
errors of CJT. Of course, the scatter around the 1-1, (solid) line
is even larger. Table~\ref{ta:comp} confirms that the rms scatter
between the results is $\pm$ 0.36 mag compared to an expected scatter
of $\pm$ 0.14 mag from our error estimates and those of CJT.  There may
also be some indication of non-linearity between the HST and
ground-based distance estimates, though more HST SBF measurements are
needed before this can be taken as established.

\begin{figure}
   \epsfig{file=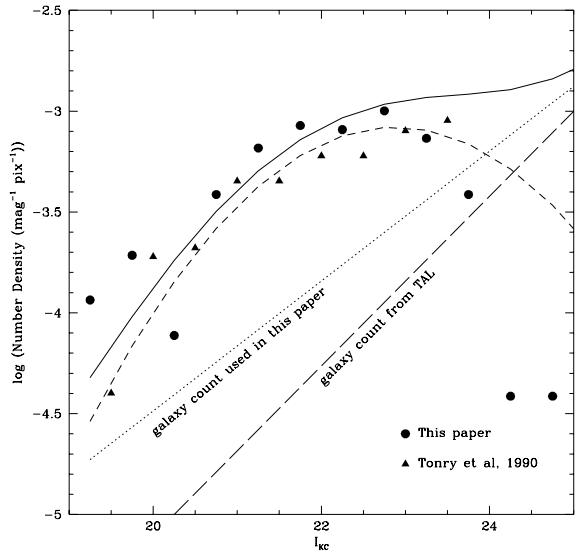,height=8.0cm}
   \caption{Point source densities in the WF2 field for NGC 4472 compared to
   those given by TAL. The values from this work have been scaled to
   0.29 arcsecond pixel size of TAL. The solid line is the fit to
   the point source number count as the sum of the galaxy number count
   (power law, shown by the dotted line) and a globular cluster
   luminosity function (a gaussian shown as a short dashed line). The
   dashed line is the galaxy number count used by TAL.}
   \label{fi:lfcomp}
\end{figure}

The most obvious candidate for the cause of  differences between the HST and
ground-based results is point source removal. With the much higher resolution of the
HST, one should do much better at point source detection and removal. However,
Figure~\ref{fi:lfcomp} shows a comparison of the density of point  source
detected in the WF2 (scaled to correct for different pixel sizes) frame of NGC
4472 compared to the point source number counts given in Tonry \etal (1990,
hereafter referred to as TAL). As can be seen, in this case at least, the
agreement between the point source counts  is very good. 

One difference with our methodology  is that, to estimate $P_r$,  TAL have taken
the galaxy count slope for the I band from Harris \shortcite{Harris88}, which
gives a value of 0.34. Although this slope is therefore steeper than the value
we have used, Figure~\ref{fi:lfcomp} actually shows that, in the case of
NGC4472, their galaxy number count lies significantly below that used
here because our zero-point offset is higher than CJT's. If CJT use our galaxy
counts their values of $P_r$ will be increased. Therefore, if the same situation
applies for for NGC 3379,  the discrepancy there could be due to CJT using an
underestimate of the galaxy counts within the field. Unfortunately,  for NGC
3379 and NGC 4406 it is not possible to determine  whether the same arguments
apply because CJT's point source count fits in these cases are not published. 

The only way we can reduce our distance for NGC3379 to improve agreement with
CJT is by {\bf not} subtracting from $P_0$ any background point  source contribution
to the variance (\ie set $P_r = 0$) , but even this does not eliminate the  large
difference  --- at most it decreases our values of $(m-M)_0$ by 0.06 mag.
Therefore, the discrepancy between CJT and ourselves cannot be due to any
overestimate we have made of the background counts, since even setting $P_r$=0 only
reduces the HST NGC3379 distance modulus by a negligible amount.

\subsection{Effects of the new $M_I:(V-I)$ calibration}

In Tonry \etal (1997) a quite different calibration of $M_I$ with
$(V-I)$ is given as
\begin{equation}
   \overline{M}_I = -1.74 + 4.5((V-I)_0 - 1.15)
\end{equation}

Taking into account the effects of absorption this implies that the distance modulus
is given by

\begin{equation}
   (m-M)_0 = m_{fluc} + 1.74 - 4.5((V-I)_{obs}-1.15) + 1.39A_B
\end{equation}

 The results of the new calibration are given in Table~\ref{ta:comp}  and shown
graphically in Figure~\ref{fi:us_tonry}. With this new calibration the mean 
difference between our results and those of CJT is only slightly reduced to 0.22
mag from 0.25 mag because only the zero-point is changed. There is also no
significant improvement in the scatter between our and CJT's results with the
r.m.s. error now $\pm$ 0.35 mag for an individual galaxy, compared to an expected
error  of $\pm$ 0.25 mag (see Table~\ref{ta:comp}).

\section {Comparison with HST SBF Results of Ajhar et al 1997}

After this paper was submitted, Ajhar \etal (1997)  published HST SBF results
for 16 galaxies, including the 3 galaxies discussed here using the same HST
data. It is therefore of interest to compare their results with ours. Ajhar
\etal only use the data from the Planetary Camera to derive their results and
in   Table~\ref{ta:comppc} we compare the fluctuation magnitudes, V-I colours
and distance moduli. Since Ajhar \etal calibrate their HST SBF distances
against the re-calibrated ground-based distances of Tonry \etal (1997), the  
new distance moduli for these 3 galaxies are 0.13$\pm$0.07 mag bigger than
for CJT, reflecting the increase in the Tonry \etal (1997) distance modulus
calibration over that of CJT.

\begin{table}  
   \begin{tabular} {|c|c|c|r|}                                               \hline
      Galaxy   & PC $m_{fluc}$    & PC $m_{fluc}$     &\multicolumn{1}{c} {diff}\\ 
               & (this paper)     &  (Ajhar \etal)    &                   \\ \hline
      NGC 3379 & $29.25 \pm 0.10$ & $29.04 \pm 0.05$  & $ 0.21 \pm 0.11$  \\
      NGC 4472 & $30.02 \pm 0.10$ & $30.09 \pm 0.04$  & $-0.07 \pm 0.11$  \\
      NGC 4406 & $29.95 \pm 0.06$ & $30.06 \pm 0.05$  & $-0.11 \pm 0.08$  \\ 
      \multicolumn{3}{c}{Mean offset}                 & $ 0.01 \pm 0.10$  \\ 
      \multicolumn{3}{c}{r.m.s. error}                & $      \pm\, 0.17$  \\ \hline
      Galaxy   & PC $(V-I)_{obs}$ & PC $(V-I)_{obs}$  &\multicolumn{1}{c} {diff} \\ 
               &    (this paper)  & (Ajhar \etal)     &                   \\ \hline
      NGC 3379 & $1.24 \pm 0.05$  & $1.260 \pm 0.006$ & $-0.020 \pm 0.05$ \\
      NGC 4472 & $1.26 \pm 0.05$  & $1.292 \pm 0.012$ & $-0.032 \pm 0.05$ \\
      NGC 4406 & $1.27 \pm 0.05$  & $1.255 \pm 0.011$ & $ 0.015 \pm 0.05$ \\  
      \multicolumn{3}{c}{Mean offset}                 & $-0.01 \pm 0.014$ \\ 
      \multicolumn{3}{c}{r.m.s. error}                & $      \pm\, 0.024$ \\ \hline
      Galaxy   & PC $(m-M)_0$     & PC $(m-M)_0$      &\multicolumn{1}{c} {diff}\\ 
               &    (this paper)  & (Ajhar \etal)     &                   \\ \hline
      NGC 3379 & $30.46 \pm 0.34$ & $30.13 \pm 0.063$ & $ 0.33 \pm 0.35$  \\
      NGC 4472 & $31.04 \pm 0.34$ & $30.90 \pm 0.088$ & $ 0.14 \pm 0.35$  \\
      NGC 4406 & $31.05 \pm 0.33$ & $31.26 \pm 0.094$ & $-0.21 \pm 0.34$  \\  
      \multicolumn{3}{c}{Mean offset}                 & $ 0.09 \pm 0.16$  \\ 
      \multicolumn{3}{c}{r.m.s. error}                & $      \pm\, 0.27$  \\ \hline
   \end{tabular}
   \caption{A comparison of the observed PC fluctuation magnitudes, V-I colours
    and distance moduli from our analysis and from the analysis of
   Ajhar \etal (1997) using the same HST data. 0.1mag has now been added to our
   fluctuation magnitudes and distance moduli to make our HST zeropoint
   consistent with that of Ajhar \etal. The final three rows include our values
   for $(m-M)_0$ assuming the  HST calibration of $M_I$ with (V-I) given by
   Ajhar \etal (1997) with a V-I coefficient of -6.5.}
   \label{ta:comppc}
\end{table}

Now it should first be noted that for the same PC data the two sets of
results are closely similar. In particular, for NGC3379 the fluctuation
magnitude of Ajhar \etal has increased to lie within 0.21 mag of our PC
value. In the mean, Table~\ref{ta:comppc} shows that the
fluctuation magnitude difference is  only 0.01$\pm$0.10 (0.17) mag and the
observed V-I colour difference is only -0.01 $\pm$ 0.014 (0.024) mag. The figures
in brackets represent the estimated r.m.s error on each individual
difference.  Dividing these by $\sqrt 2$, suggests that the error on each
fluctuation magnitude is $\pm0.12$ mag and on each $V-I$ colour is
$\pm$0.017 mag.  The error on an individual difference in $(m-M)_0$ is
$\pm$0.27 mag with the error being approximately equally distributed between
the colour and the fluctuation magnitude errors. Thus although the
systematic differences between the two analyses are small, the errors on
individual measurements are non-negligible at $\pm$0.12 mag and $\pm$0.19 mag
on individual measurement of $m_{fluc}$ and $(m-M)_0$ and closer to our
error estimates than those of Ajhar \etal. However, the level of
disagreement in measuring $m_{fluc}$ at $\pm$0.17 mag is too small to explain
the large, $\pm$0.25 mag, scatter seen in the previous comparison with the
results of CJT in  Table~\ref{ta:comp}.

Table~\ref{ta:compmm} further compares  the HST distance moduli  of
Ajhar \etal (1997) and our results, these also now including the
three WF camera frames, as well as the PC frame. Table~\ref{ta:compmm} also
intercompares our results with the new ground-based SBF distance moduli of
Tonry \etal (1997), published for the first time by Ajhar \etal (1997).  Taking
the top two sets of rows first, where our results now assume the new, steep  
slope (-6.5),  HST $M_I:V-I$ calibration of Ajhar \etal with  we first note
that although the systematic errors in the mean remain insignificant, the rms
errors in the individual $(m-M)_0$  differences have now increased to
$\pm$0.42 mag in the case of the Ajhar \etal data and to $\pm$0.43 mag in the
case of the Tonry \etal data. In the case of NGC3379, the previous 0.33 mag
discrepancy in the PC \DM  has increased to 0.65 mag. This is because our
overall fluctuation magnitude has increased by a further 0.16 mag and this has
not been offset by any increase in V-I  colour; indeed the reverse is true
particularly in the case of chip WF4 (see Table~\ref{ta:comp}.) 

In general, there is no evidence in the correlation between our results for
the  four HST frames that the new steep HST $M_I:V-I$ calibration with slope
-6.5 (Ajhar \etal 1997) leads to a smaller dispersion in distance; indeed there
is clear evidence that it increases it (see Table~\ref{ta:compmm}.), since our
errors  in $(m-M)_o$ significantly increase when this relation is used rather
than that of CJT (see bottom  two sets of rows in Table~\ref{ta:compmm}). It is
unlikely that this is just due to our own measurement errors since the
agreement in V-I between ourselves and Ajhar \etal  for the PC V-I colour is
good with the error being -0.012$\pm$0.014 (0.024) mag and this is generally
much smaller than the dispersion in colour between the four WFPC2 CCDs
($\pm$0.054 mag for NGC3379, $\pm$0.049 mag for NGC4406, $\pm$0.015 mag for
NGC4472). The suggestion is that either the intrinsic V-I measurement errors in
the HST data are much larger than claimed by Ajhar \etal or the correlation
between $m_{fluc}$ and colour for the HST bands is much less steep  than they
claim. 

 \begin{table}
   \begin{tabular} {|c|c|c|r|}                                               \hline
      Galaxy   & $(m-M)_0$        & $(m-M)_0$          &\multicolumn{1}{c} {diff}\\ 
               & (this/HST)       &(Ajhar \etal 97)    &                  \\ \hline
      NGC 3379 & $30.78 \pm 0.33$ & $ 30.13 \pm 0.063$ & $ 0.65 \pm 0.34$ \\
      NGC 4472 & $30.95 \pm 0.33$ & $ 30.90 \pm 0.088$ & $ 0.05 \pm 0.34$ \\
      NGC 4406 & $31.11 \pm 0.33$ & $ 31.26 \pm 0.094$ & $-0.15 \pm 0.34$ \\
      \multicolumn{3}{c}{Mean offsets}                 & $ 0.18 \pm 0.24$ \\
      \multicolumn{3}{c}{r.m.s. error}                 & $      \pm\, 0.42$ \\ \hline
      Galaxy   & $(m-M)_0$        & $(m-M)_0$          &\multicolumn{1}{c} {diff}\\ 
               & (this/HST)       & (Tonry \etal 97)   &                  \\ \hline
      NGC 3379 & $30.78 \pm 0.33$ & $ 30.08 \pm 0.080$ & $ 0.70 \pm 0.34$ \\
      NGC 4472 & $30.95 \pm 0.33$ & $ 30.94 \pm 0.068$ & $ 0.01 \pm 0.34$ \\
      NGC 4406 & $31.11 \pm 0.33$ & $ 31.19 \pm 0.070$ & $-0.08 \pm 0.34$ \\
      \multicolumn{3}{c}{Mean offsets}                 & $ 0.21 \pm 0.25$ \\
      \multicolumn{3}{c}{r.m.s. error}                 & $      \pm\, 0.43$ \\ \hline
      Galaxy   & $(m-M)_0 $       & $(m-M)_0$          &\multicolumn{1}{c} {diff}\\ 
               & (this/CJT)       & (Ajhar \etal 97)   &                  \\ \hline
      NGC 3379 & $30.50 \pm 0.11$ & $ 30.13 \pm 0.063$ & $ 0.37 \pm 0.13$ \\
      NGC 4472 & $30.97 \pm 0.12$ & $ 30.90 \pm 0.088$ & $ 0.07 \pm 0.15$ \\
      NGC 4406 & $31.16 \pm 0.11$ & $ 31.26 \pm 0.094$ & $-0.10 \pm 0.14$ \\
      \multicolumn{3}{c}{Mean offsets}                 & $ 0.11 \pm 0.14$ \\ 
      \multicolumn{3}{c}{r.m.s. error}                 & $      \pm\, 0.24$ \\ \hline
      Galaxy   & $(m-M)_0$        & $(m-M)_0$          &\multicolumn{1}{c} {diff}\\
               & (this/CJT)       & (Tonry \etal 97)   &                  \\ \hline
      NGC 3379 & $30.50 \pm 0.11$ & $ 30.08 \pm 0.080$ & $ 0.42 \pm 0.14$ \\
      NGC 4472 & $30.97 \pm 0.12$ & $ 30.94 \pm 0.068$ & $ 0.03 \pm 0.14$ \\
      NGC 4406 & $31.16 \pm 0.11$ & $ 31.19 \pm 0.070$ & $-0.03 \pm 0.13$ \\
      \multicolumn{3}{c}{Mean offsets}                 & $ 0.14 \pm 0.14$ \\ 
      \multicolumn{3}{c}{r.m.s. error}                 & $      \pm\, 0.24$ \\ \hline

   \end{tabular}
   \caption{ A comparison our HST (WF+PC) distance moduli calibrated using 
   either the HST $M_I:V-I$ calibration of Ajhar \etal (1997) or the original
   calibration of CJT as indicated, with the HST results of Ajhar \etal
   and the ground-based results of Tonry \etal (1997).  Again, in the
   top two  sets of rows where our results assume the HST calibration of
   Ajhar \etal, we have increased our distance moduli by 0.1 mag to be
   consistent with their HST zeropointing procedure.} 
   \label{ta:compmm}
\end{table}

The comparison of our 4-frame  result with the 1-frame result of Ajhar
\etal in the third set of rows in Table~\ref{ta:compmm} also shows that the 
r.m.s. differences at $\pm$0.24 mag for each galaxy are reduced from those
found in our comparison with the previous ground-based results of CJT (
$\pm$0.36 mag). Splitting this error equally between Ajhar \etal and
ourselves would give an estimated $\pm$0.17 mag error on an individual HST
SBF measurement. However, the r.m.s. difference at $\pm$0.24 mag is still
significantly bigger than indicated by combining our formal SBF 
measurement error at  $\sim\pm$0.12 mag with that of  Ajhar \etal at
$\sim\pm$0.08 mag. The comparison of our HST SBF moduli with the new
ground-based moduli of Tonry \etal 1997  in the final  set of rows in
Table~\ref{ta:compmm} leads to similar conclusions.

Ajhar \etal (1997) also  concluded that  measurement errors alone were not
enough to explain the dispersion between their new ground-based and HST
observations. However, the final average, {\it total}  error  of $\pm$0.12
mag quoted by them for  their HST SBF distance moduli is still lower than
we derive.   If the galaxy NGC4621, classed as an `outlier', is
re-included in their sample, then their total error on each HST distance
modulus rises  to $\pm$0.17 mag which is  close to what we estimate.

Thus, overall, the analyses of Ajhar \etal  and ourselves seem to show
reasonable agreement in $m_{fluc}$ and V-I measurements in the PC images
of these three galaxies. They also verify that the measurement errors on
these quantities are consistent with our estimates. However, significant
disagreements remain between our overall HST SBF results and the HST and
ground-based SBF results of Ajhar \etal which suggest that the errors on
the SBF method may be significantly larger than  expected from simply
combining  measurement errors.

\section{Conclusions}

We have discovered a  disagreement between  HST SBF distances and those
previously determined by CJT using ground-based images. In the case of one
galaxy, NGC 3379, the discrepancy is 0.63$\pm$0.13 mag and is significant at
the 4.8$\sigma$ level. From the overall comparison, the real error on the
ground-based SBF distance moduli of CJT could be as high  as $\pm$0.25 mag on
average, compared to the  $\pm$0.08 mag claimed by CJT from  measurement
errors. A possible cause of the disagreement in the case of NGC3379 is that
the error associated with the background galaxy correction has been
underestimated by CJT; our HST results are insensitive to this type of error. 

Improved agreement is seen with   the HST SBF  results of Ajhar \etal (1997)
and the new ground-based results of Tonry \etal (1997).  In particular,  on
the same individual HST PC frame, the results of Ajhar \etal and ourselves 
for $m_{fluc}$ agree to $\pm$0.17mag and for V-I to $\pm$0.017mag. In the
case of the distance modulus, the comparison of our 4-frame HST results with
both the above HST and ground-based results  suggest that the error is
probably reduced to $\pm$0.17mag for an individual  galaxy distance modulus.
However, this is still significantly larger than claimed by these authors.
Also, we find no evidence  in our WFPC2 data to support the steep slope of
-6.5  for the HST  $M_I:V-I$ relation advocated by Ajhar \etal (1997).
Further, since the HST data of Ajhar \etal is a subset of that used here, 
these two results are not independent and the real error on HST SBF distance
estimates could be larger. Finally, it should be noted that the error
estimates presented here form a lower limit to the true SBF error because all
points of comparison accept the precepts of the SBF method and make no
reference to independent methods of distance estimation.

\end{document}